\documentclass[prd,twocolumn,groupedaddress,showpacs,nofootinbib]{revtex4}
\usepackage{graphicx}
\usepackage{dcolumn}
\usepackage{amssymb}
\usepackage{mathrsfs}
\usepackage{amsmath}
\usepackage{epsfig}
\usepackage[dvips]{color}
\usepackage{hhline}

\begin{document}

\title{Common origin of baryon asymmetry and proton decay}

\author{Pei-Hong Gu$^{1}_{}$}
\email{peihong.gu@mpi-hd.mpg.de}

\author{Utpal Sarkar${}^{2,3}$}
\email{utpal@prl.res.in}

\affiliation{ ${}^1$Max-Planck-Institut f\"{u}r Kernphysik,
Saupfercheckweg 1,
69117 Heidelberg, Germany\\
${}^{2}$Physical Research Laboratory, Ahmedabad 380009, India\\
${}^3$McDonnell Center for the Space Sciences, Washington
University, St. Louis, MO 63130, USA}

\begin{abstract}

A successful baryogenesis theory requires a baryon-minus-lepton
number violation if it works before the electroweak phase
transition. The baryon-minus-lepton number violation could also
exist in some proton decay modes. We propose a model to show that
the cosmological baryon asymmetry and the proton decay could have a
common origin. Specifically, we introduce an isotriplet and two
isosinglet leptoquark scalars as well as two isotriplet Higgs
scalars to the canonical seesaw model. The decays of the Higgs
triplets can generate a desired baryon-minus-lepton asymmetry in the
leptoquarks. After the Higgs triplets pick up their
seesaw-suppressed vacuum expectation values, the leptoquarks with
TeV-scale masses can mediate a testable proton decay.

\end{abstract}


\maketitle

\section{Introduction}

There is an $SU(2)_L^{}$ global anomaly \cite{thooft1976} violating
the baryon $(B)$ and lepton $(L)$ numbers by an equal amount in the
$SU(3)_c^{}\times SU(2)_L^{}\times U(1)_Y^{}$ standard model (SM).
At the finite temperatures $100\,\textrm{GeV} \lesssim T \lesssim
10^{12}_{}\,\textrm{GeV}$, the anomalous process becomes strong due
to an instanton-like solution, the so-called sphalerons
\cite{krs1985}. During the sphaleron epoch, neither the baryon
asymmetry nor the lepton asymmetry can survive if the baryon and
lepton asymmetries are equal. However, the sphaleron processes will
not affect any primordial $B-L$ asymmetry and will convert the $B-L$
asymmetry to a baryon asymmetry and a lepton asymmetry
\cite{fy1986}. So, a successful baryogenesis mechanism working above
the weak scale should require a $B-L$ number violation which is a
pure baryon number violation, a pure lepton number violation or a
combined baryon and lepton violation.

The baryon and/or lepton number violation can lead to other
interesting phenomena. For example, we can obtain a Majorana
neutrino mass term by a lepton number violation of two units, a
neutron-antineutron oscillation by a baryon number violation of two
units, as well as a two-body proton decay by a baryon number
violation of one unit and a lepton number violation of one unit.

In the simplest grand unified theories (GUTs), a baryon asymmetry
and an equal lepton asymmetry can be simultaneously produced at the
GUT scale through some baryon and lepton number violating
interactions, which are also responsible for generating a $B-L$
conserving proton decay. In this GUT baryogenesis scenario, the
baryon and lepton asymmetries will be both wiped out by the
sphaleron processes.

In this paper we shall show it is possible to realize the
baryogenesis and the proton decay by same interactions. For this
purpose, we shall extend the SM by an isotriplet and two isosinglet
leptoquark scalars, two isotriplet Higgs scalars as well as three
right-handed neutrinos. The Majorana masses of the right-handed
neutrinos will softly break the lepton number while the trilinear
scalar couplings involving the Higgs triplets will softly break both
of the baryon and lepton numbers. The $B-L$ number violating
processes involving the right-handed neutrinos will be assumed to
decouple before the out-of-equilibrium decays of the Higgs triplets.
So, the $B-L$ asymmetry from the decays of the Higgs triplets into
the leptoquarks can explain the baryon asymmetry in the universe.
After the Higgs triplets pick up their seesaw-suppressed vacuum
expectation values (VEVs), the leptoquarks can mediate a $B-L$
violating decay \cite{pss1983} of the proton into two antineutrinos
and one positron or antimuon. If the leptoquarks are at the TeV
scale, the proton decay will be close to the experimental limit.

\section{The model}

For simplicity, we do not write down the full lagrangian. Instead,
we only give the terms as below,
\begin{eqnarray}
\label{lagrangian}
\mathcal{L}&\supset&-y_{ik}^{}\bar{l}_{L_i^{}}^{}\phi
N_{R_k^{}}^{}-\frac{1}{2}M_{N_{k}^{}}^{}\bar{N}_{R_k^{}}^c
N_{R_k^{}}^{}
-f_{ij}^{\Omega}\bar{l}_{L_i^{}}^c i\tau_2^{}\Omega q_{L_j^{}}^{} \nonumber\\
&&-f_{ij}^{\delta_a^{}}\delta_a^{}\bar{l}_{L_i^{}}^c i\tau_2^{}
q_{L_j^{}}^{}-h_{ij}^{\delta_a^{}}\delta_a^{}\bar{e}_{R_i^{}}^c
u_{R_{j}^{}}^{}-\mu_a^{}\phi^T_{}i\tau_2^{}\Sigma_a^{}
\phi\nonumber\\
&&-\kappa_a^{}\delta_1^{}\delta_2^{}\textrm{Tr}\left(\Sigma^\dagger_a\Omega\right)
 +\textrm{H.c.}\nonumber\\
&=&-y_{ik}^{}(\bar{\nu}_{L_i^{}}^{}\phi^0_{}+\bar{e}_{L_i^{}}^{}\phi^{-}_{})N_{R_k^{}}^{}
-\frac{1}{2}M_{N_{k}^{}}^{}\bar{N}_{R_k^{}}^c
N_{R_k^{}}^{}\nonumber\\
&&-f_{ij}^{\Omega}\left[\omega^{-2/3}\bar{\nu}_{L_i^{}}^c
u_{L_j^{}}^{}-\frac{1}{\sqrt{2}}\omega^{+1/3}_{}(\bar{\nu}_{L_i^{}}^c
d_{L_j^{}}^{}+\bar{e}_{L_i^{}}^c
u_{L_j^{}}^{} )\right.\nonumber\\
&& \left.- \omega^{+4/3}_{}\bar{e}_{L_i^{}}^c
d_{L_j^{}}^{}\right]-f_{ij}^{\delta_a^{}}\delta_a^{}(\bar{\nu}_{L_i^{}}^c
d_{L_j^{}}^{}-\bar{e}_{L_i^{}}^c
u_{L_j^{}}^{})\nonumber\\
&&-h_{ij}^{\delta_a^{}}\delta_a^{}\bar{e}^c_{R_i^{}}u_{R_j^{}}^{}-\mu_a^{}(\sigma^0_{a}\phi^0_{}\phi^0_{}
+\sqrt{2}\sigma^{+}_{a}\phi^0_{}\phi^{-}_{}\nonumber\\
&&
-\sigma^{++}_{a}\phi^{-}_{}\phi^{-}_{})-\kappa_a^{}\delta^{+1/3}_1\delta^{+1/3}_2
\left[(\sigma^{0}_{a})^\ast_{}\omega^{-2/3}_{}
\right.\nonumber\\
&&
\left.+(\sigma^{+}_{a})^\ast_{}\omega^{+1/3}_{}+(\sigma^{++}_{a})^\ast_{}\omega^{+4/3}_{}\right]+\textrm{H.c.}\,,
\end{eqnarray}
with the SM quarks, leptons and Higgs scalar:
\begin{eqnarray}
q_{L_i^{}}^{}(\textbf{3},\textbf{2},+\frac{1}{6})&=&\left[\begin{array}{l}u_{L_i^{}}^{}\\
[1.5mm]d_{L_i^{}}^{}\end{array}\right]\,,~u_{R_i^{}}^{}(\textbf{3},\textbf{1},+\frac{2}{3})\,,\nonumber\\
[3mm]
l_{L_i^{}}^{}(\textbf{1},\textbf{2},-\frac{1}{2})&=&\left[\begin{array}{l}\nu_{L_i^{}}^{}\\
[1.5mm]e_{L_i^{}}^{}\end{array}\right]\,,~e_{R_i^{}}^{}(\textbf{1},\textbf{1},-1)\,,\nonumber\\
[3mm]
\phi(\textbf{1},\textbf{2},-\frac{1}{2})&=&\left[\begin{array}{l}\phi^0_{}\\
[1.5mm]\phi^{-}_{}\end{array}\right]\,,
\end{eqnarray}
the gauge-singlet right-handed neutrinos:
\begin{eqnarray}
N_{R_i^{}}^{}(\textbf{1},\textbf{1},0)\,,
\end{eqnarray}
the $[SU(2)]$-singlet and triplet leptoquark scalars:
\begin{eqnarray}
&&\delta_a^{}(\textbf{3},\textbf{1},\frac{1}{3})=\delta^{+1/3}_{a}\,,\nonumber\\
&&\Omega(\textbf{3},\textbf{3},\frac{1}{3})=
\left[\begin{array}{rr} \frac{1}{\sqrt{2}}\omega^{+1/3}_{} &~~ \omega^{+4/3}_{}\\
[3mm] \omega^{-2/3}_{} & ~~-\frac{1}{\sqrt{2}}\omega^{+1/3}_{}
\end{array}\right]\,,
\end{eqnarray}
and the $[SU(2)]$-triplet Higgs scalars:
\begin{eqnarray}
 \Sigma_a^{}(\textbf{1},\textbf{3},1)&=&
 \left[\begin{array}{rr} \frac{1}{\sqrt{2}}\sigma^{+}_{a} & ~~\sigma^{++}_{a}\\
[3mm] \sigma^{0\,}_{a} & ~~-\frac{1}{\sqrt{2}}\sigma^{+~~}_{a}
\end{array}\right]\,.
\end{eqnarray}

We assign the baryon and lepton numbers as below,
\begin{eqnarray}
(B,L)=\left\{\begin{array}{l}(\frac{1}{3},0)~\textrm{for}~
q_L^{}~\textrm{and~other~SM~quarks}\,,\\
[3mm] (0,+1)~\textrm{for}~
N_R^{}\,,~l_L^{}~\textrm{and~other~SM~leptons}\,,\\
[3mm]
(-\frac{1}{3},-1)~\textrm{for}~ \Omega~\textrm{and}~\delta\,,\\
[3mm]
(-1,-3)~\textrm{for}~ \Sigma\,,\\
[3mm] (0,0)~\textrm{for}~ \phi\,.\end{array}\right.\nonumber
\end{eqnarray}
\vspace{-2.4cm}
\begin{eqnarray}
\end{eqnarray}
\vspace{0.35cm}
\begin{eqnarray}
\nonumber
\end{eqnarray}
In Eq. (\ref{lagrangian}), the Majorana masses of the right-handed
neutrinos break the lepton number, the trilinear couplings of the
Higgs scalars break both the baryon number and the lepton number,
while other terms conserve the baryon and lepton numbers. Clearly,
the Majorana masses and the trilinear scalar couplings also break
the $B-L$ number. Note that the baryon and lepton numbers are only
allowed to softly break. We hence have forbidden the Yukawa
couplings of the Higgs triplets to the leptons and the Yukawa
couplings of the leptoquarks to the quarks.

The Higgs doublet $\phi$ will develop a VEV:
\begin{eqnarray}
\langle\phi^0_{}\rangle\simeq 174\,\textrm{GeV}\,,
\end{eqnarray}
to spontaneously break the electroweak symmetry. The Higgs triplets
$\Sigma_a^{}$ then can pick up their seesaw-suppressed VEVs:
\begin{eqnarray}
\langle\sigma^0_a\rangle\simeq
-\frac{\mu_a^\ast\langle\phi^0_{}\rangle^2_{}}{M_{\Sigma_a}^2}\ll\langle\phi^0_{}\rangle\,,
\end{eqnarray}
like the Higgs triplets in the type-II seesaw model \cite{mw1980}.

\section{Neutrino masses}

\begin{figure*}
\vspace{7.5cm} \epsfig{file=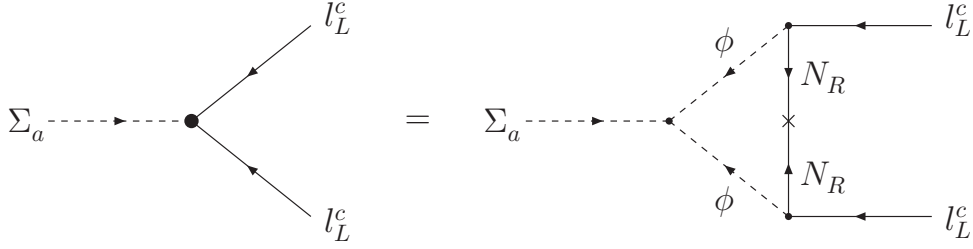, bbllx=6cm, bblly=6.0cm,
bburx=16cm, bbury=16cm, width=9cm, height=9cm, angle=0, clip=0}
\vspace{-12.7cm} \caption{\label{tyukawa} The effective Yukawa
couplings of the Higgs triplets to the lepton doublets. The CP
conjugation is not shown for simplicity.}
\end{figure*}

Although the Yukawa couplings of the Higgs triplets to the lepton
doublets are absent from Eq. (\ref{lagrangian}), they can appear at
one-loop order as shown in Fig. \ref{tyukawa}. We calculate the
effective Yukawa couplings to be
\begin{eqnarray}
\label{tbyukawa}
\mathcal{L}&\supset&-\frac{1}{2}(g_a^{})_{ij}^{}\bar{l}_{L_i}^c
i\tau_2^{}\Sigma_a^{} l_{L_j}^{}+\textrm{H.c.}\nonumber\\
&=&-\frac{1}{2}(g_a^{})_{ij}^{}(\sigma_a^0\bar{\nu}_{L_i}^c
\nu_{L_j}^{}+\sqrt{2}\sigma_a^{+}\bar{\nu}_{L_i}^c
e_{L_j}^{}-\sigma_a^{++}\bar{e}_{L_i}^c
e_{L_j}^{})\nonumber\\
&&+\textrm{H.c.}\,.
\end{eqnarray}
with
\begin{eqnarray}
(g_a^{})_{ij}^{}&\simeq&\frac{1}{4\pi^2_{}}\mu_a^{}\left(-\frac{M_{\Sigma_a}^2}{M_{N_k}^2}+i\pi\right)
y_{ik}^\ast\frac{1}{M_{N_k}^{}}y_{jk}^\ast\nonumber\\
&\simeq&
\frac{i}{4\pi}\frac{\mu_a^{}(m_{\nu}^{})_{ij}^{}}{\langle\phi^0_{}\rangle^2_{}}
~~\textrm{for}~~\frac{M_{\Sigma_a}^2}{M_{N_k}^2}\ll 1\,.
\end{eqnarray}
Here the canonical type-I seesaw \cite{minkowski1977} formula have
been adopted,
\begin{eqnarray}
\mathcal{L}\supset-\frac{1}{2}m_\nu^{}\bar{\nu}_L^c \nu_L^{}
 +\textrm{H.c.}~~\textrm{with}~~
m_\nu^{}=-y^\ast_{}\frac{\langle\phi^0_{}\rangle^2_{}}{M_N^{}}y^\dagger_{}\,.
\end{eqnarray}
The effective Yukawa couplings (\ref{tbyukawa}) will also contribute
to the neutrino masses through the type-II seesaw mechanism,
\begin{eqnarray}
(\delta
m_\nu^{})_{ij}^{}=(g_a^{})_{ij}^{}\sum_a^{}\langle\sigma^0_a\rangle
\simeq-
\frac{i}{4\pi}\sum_a^{}\frac{|\mu_a^{}|^2_{}}{M_{\Sigma_a}^2}(m_\nu^{})_{ij}^{}
\,.
\end{eqnarray}
Clearly, the type-I seesaw could dominate over the type-II seesaw,
\begin{eqnarray}
\delta m_\nu^{}\ll m_\nu^{}~~\textrm{for}~~|\mu_a^{}|^2_{}\ll
M_{\Sigma_a}^2 \,.
\end{eqnarray}

The neutrino mass matrix can be diagonalized by
\begin{eqnarray}
m_\nu^{}=U \textrm{diag}\{m_1^{},m_2^{},m_3^{}\}U^T_{}\,,
\end{eqnarray}
where $m_{1,2,3}^{}$ are the mass eigenvalues while $U$ is the
mixing matrix with three mixing angles, one Dirac CP phase and two
Majorana CP phases. The neutrino oscillation experiments have given
some information on the neutrino masses and mixing such as
\cite{stv2011}
\begin{subequations}
\begin{eqnarray}
m_2^2-m_1^2&=&(7.09-8.19)\times 10^{-5}_{}\,\textrm{eV}^2_{}\,;\\
m_3^2-m_1^2&=&\left\{\begin{array}{r}-(2.08-2.64)\times 10^{-3}_{}\,\textrm{eV}^2_{}\,,\\
(2.18-2.73)\times 10^{-3}_{}\,\textrm{eV}^2_{}\,.\end{array}\right.
\end{eqnarray}
\end{subequations}
Furthermore, the cosmological observations \cite{komatsu2011} have
put an upper bound on the sum of the neutrino mass eigenvalues,
\begin{eqnarray}
\sum_i^{}m_i^{}&<&0.58\,\textrm{eV}\,.
\end{eqnarray}

\section{Baryogenesis}

We assume that the lepton number violating processes involving the
right-handed neutrinos will decouple before the decays of the Higgs
triplets and then the final $B-L$ asymmetry should be generated by
the decays of the Higgs triplets. Below the seesaw scale $M_N^{}$,
the $\Delta L =2 $ scattering processes should have the rate
\cite{fy1990}:
\begin{eqnarray}
\Gamma_A^{}=\frac{1}{\pi^{3}_{}}\frac{\sum
m_i^2}{\langle\phi^0_{}\rangle^4_{}}T^{3}_{}\,.
\end{eqnarray}
By requiring
\begin{eqnarray}
\Gamma_A^{} < H(T)\,,
\end{eqnarray}
where the Hubble constant is given by
\begin{eqnarray}
H(T)=\left(\frac{8\pi^{3}_{}g_{\ast}^{}}{90}\right)^{\frac{1}{2}}_{}
\frac{T^{2}_{}}{M_{\textrm{Pl}}^{}}\,,
\end{eqnarray}
with $M_{\textrm{Pl}}^{}=1.22\times 10^{19}_{}\,\textrm{GeV}$ being
the Planck mass and $g_\ast^{}=106.75+30=136.75$ being the
relativistic degrees of freedom (the SM fields plus an isotriplet
and two isosinglet leptoquark scalars), the lepton number violating
processes will decouple when the temperature falls down to
\begin{eqnarray}
T\simeq 2\times 10^{13}_{}\,\textrm{GeV} \left(\frac{2.5\times
10^{-3}_{}\,\textrm{eV}^2_{}}{\sum_{i}^{} m_i^2}\right) \,.
\end{eqnarray}
In the following demonstration, we hence shall consider the mass
spectrum as below,
\begin{eqnarray}
M_{\Sigma_{1,2}}^{}< 10^{13}_{}\,\textrm{GeV} < M_N^{}\,.
\end{eqnarray}

\begin{figure*}
\vspace{4.7cm} \epsfig{file=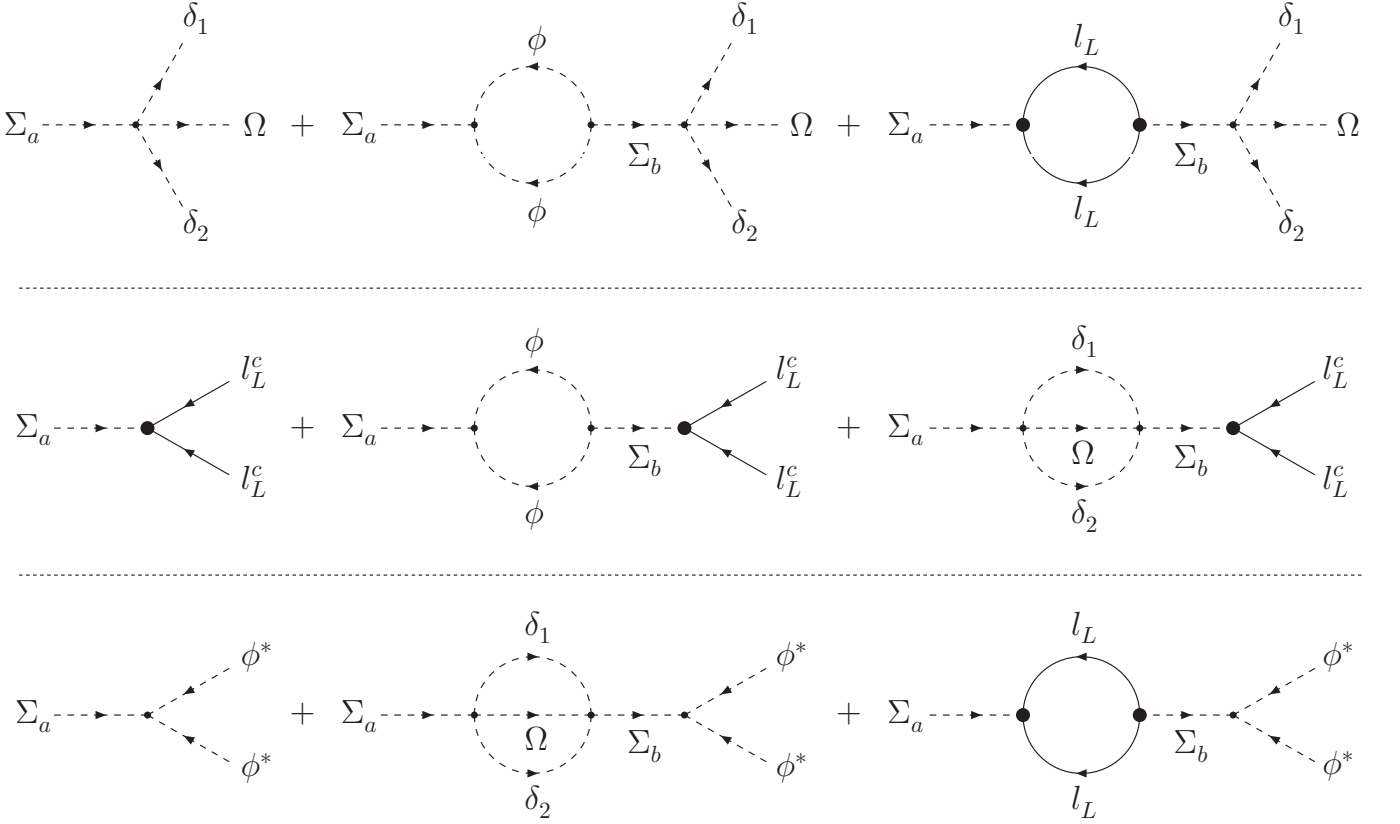, bbllx=5.7cm, bblly=6.0cm,
bburx=15.7cm, bbury=16cm, width=8.8cm, height=8.8cm, angle=0,
clip=0} \vspace{-2cm} \caption{\label{baryogenesis} The decays of
the Higgs triplets. The CP conjugation is not shown for simplicity.}
\end{figure*}

As shown in Fig. \ref{baryogenesis}, the Higgs triplets have the
following decay modes:
\begin{eqnarray}
\left\{\begin{array}{lclll}\Sigma
&\rightarrow&\delta_1^{}~\,\delta_2^{}~\,\Omega\,,&\phi^\ast_{}\phi^\ast_{}\,,&l_L^c\, l_L^c\,;\\
[3mm] \Sigma^\ast_{}
&\rightarrow&\delta^\ast_{1}~\delta^\ast_{2}~\,\Omega^\ast_{}\,,&\phi~\,\phi\,,~&l_L^{}\,l_L^{}\,.\end{array}\right.
\end{eqnarray}
Therefore, the decays of the Higgs triplets can produce a $B-L$
asymmetry in the leptoquarks and the leptons if CP is not conserved.
To quantify the $B-L$ asymmetry, we can define a CP asymmetry in the
decays of the Higgs triplets $\Sigma_a^{}$,
\begin{eqnarray}
\varepsilon_{a}^{}=\varepsilon_{a}^{\delta_1^{}\delta_2^{}\Omega}+\varepsilon_{a}^{l_L^{}}\,,
\end{eqnarray}
with
\begin{subequations}
\begin{eqnarray}
\varepsilon_{a}^{\delta_1^{}\delta_2^{}\Omega}&=&2\frac{\Gamma_{\Sigma_a^{}
\rightarrow\delta_1^{}\delta_2^{}\Omega}^{}-\Gamma_{\Sigma^\ast_a
\rightarrow\delta^\ast_{1}\delta^\ast_{2}\Omega^\ast_{}}^{}}{\Gamma_{a}^{}}\,,\\
\varepsilon_{a}^{l_L^{}}&=&2\frac{\Gamma_{\Sigma_a^{} \rightarrow
l_L^c l_L^c}^{}-\Gamma_{\Sigma^\ast_a \rightarrow l_L^{}
l_L^{}}}{\Gamma_{a}^{}}\,.
\end{eqnarray}
\end{subequations}
Here
\begin{eqnarray}
\Gamma_a^{}&=&\Gamma_{\Sigma_a^{}}^{}=\Gamma_{\Sigma_a^{}
\rightarrow\delta_1^{}\delta_2^{}\Omega}^{}+\Gamma_{\Sigma_a^{}
\rightarrow\phi^\ast_{}\phi^\ast_{}}^{}+\Gamma_{\Sigma_a^{}
\rightarrow l_L^c l_L^c}^{}\nonumber\\
[2mm] &\equiv&\Gamma_{\Sigma^\ast_a}^{}=\Gamma_{\Sigma^\ast_a
\rightarrow\delta^\ast_{1}\delta^\ast_{2}\Omega^\ast_{}}^{}+\Gamma_{\Sigma_a^\ast
\rightarrow\phi\phi}^{}+\Gamma_{\Sigma_a^\ast \rightarrow l_L^{}
l_L^{}}^{}\quad
\end{eqnarray}
is the decay width. We can calculate the tree-level decay width:
\begin{eqnarray}
\Gamma_a^{}&=&\frac{1}{8\pi}
\left(\frac{1}{64\pi^2_{}}|\kappa_a^{}|^2_{}+\frac{|\mu_a^{}|^2_{}}{M^2_{\Sigma_a}}+\frac{1}{64\pi^2_{}}
\frac{|\mu_a^{}|^2_{}\sum_i^{}m_i^2}{\langle\phi^0_{}\rangle^4}\right)\nonumber\\
&&\times M_{\Sigma_a}^{}\,,
\end{eqnarray}
and the one-loop CP asymmetries:
\begin{subequations}
\begin{eqnarray}
\varepsilon_a^{\delta_1^{}\delta_2^{}\Omega}&=& \frac{1}{2\pi}
\frac{\textrm{Im}(\kappa_a^\ast \kappa_b^{} \mu_a^{}
\mu_b^\ast)}{|\kappa_a^{}|^2_{}}\frac{1}
{M_{\Sigma_a}^2-M_{\Sigma_b}^2}\nonumber\\
&&\times
\left(1+\frac{1}{64\pi^2}\frac{M_{\Sigma_a}^2\sum_i^{}m_i^2}{\langle\phi^0_{}\rangle^4_{}}\right)
\nonumber\\
&& \times
\frac{\frac{1}{64\pi^2_{}}|\kappa_a^{}|^2_{}}{\frac{1}{64\pi^2_{}}|\kappa_a^{}|^2_{}+\frac{|\mu_a^{}|^2_{}}
{M^2_{\Sigma_a}}+\frac{1}{64\pi^2_{}}
\frac{|\mu_a^{}|^2_{}\sum_k^{}m_k^2}{\langle\phi^0_{}\rangle^4}}\,,\quad\\
\varepsilon_a^{l_L^{}}&=&-
\frac{1}{128\pi^3_{}}\frac{\textrm{Im}(\kappa_a^\ast \kappa_b^{}
\mu_a^\ast \mu_b^{})}{|\mu_a^{}|^2_{}}\frac{M_{\Sigma_a}^2}
{M_{\Sigma_a}^2-M_{\Sigma_b}^2}
\nonumber\\
&& \times \frac{\frac{1}{64\pi^2_{}}
\frac{|\mu_a^{}|^2_{}\sum_i^{}m_i^2}{\langle\phi^0_{}\rangle^4}}{\frac{1}{64\pi^2_{}}|\kappa_a^{}|^2_{}+\frac{|\mu_a^{}|^2_{}}
{M^2_{\Sigma_a}}+\frac{1}{64\pi^2_{}}
\frac{|\mu_a^{}|^2_{}\sum_k^{}m_k^2}{\langle\phi^0_{}\rangle^4}}\,.
\end{eqnarray}
\end{subequations}

When the Higgs triplets $\Sigma_a^{}$ go out of equilibrium, their
CP violating decays can generate a $B-L$ asymmetry in the
leptoquarks $\Omega$ and $\delta_{1,2}$ as well as the leptons
$l_L^{}$. For example, we consider the weak washout region, where
the out-of-equilibrium condition can be described by the following
quantity,
\begin{eqnarray}
\label{weakwashout}
K_a^{}=\frac{\Gamma_a^{}}{H}\left|_{T=M_{\Sigma_a}^{}}^{}\right.\lesssim
1\,.
\end{eqnarray}
The induced $B-L$ asymmetry then can approximate to \cite{kt1990}
\begin{eqnarray}
\label{blasymmetry} \frac{n_{B-L}^{}}{s}\sim 3\times
\frac{\varepsilon_a^{}}{g_\ast^{}}\quad \textrm{for}\quad
K_a^{}\lesssim 1\,,
\end{eqnarray}
where the factor $3$ means the three components of the decaying
Higgs triplets. If $\Sigma_1^{}(\Sigma_2^{})$ is much lighter than
$\Sigma_2^{}(\Sigma_1^{})$, the final $B-L$ asymmetry should come
from the decays of $\Sigma_1^{}(\Sigma_2^{})$. Alternatively, if
$\Sigma_1^{}$ and $\Sigma_2^{}$ have a small mass split, both of
them will significantly contribute to the final $B-L$ asymmetry. In
this case, the CP asymmetry could be resonantly enhanced
\cite{fps1995}.

Since the leptoquarks decays into the leptons and the quarks, their
$B-L$ asymmetries can be transferred to a baryon asymmetry and a
lepton asymmetry through the sphaleron processes. The final baryon
asymmetry in the universe should be \cite{krs1985}
\begin{eqnarray}
\label{basymmetry} \frac{n_B^{}}{s}=\frac{28}{79}
\frac{n_{B-L}^{}}{s}\,.
\end{eqnarray}

\section{Proton decay}

\begin{figure}
\vspace{11cm} \epsfig{file=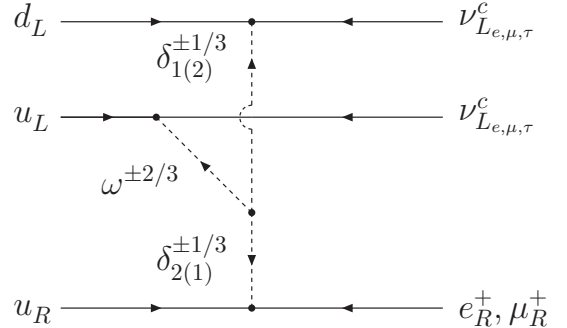, bbllx=1cm, bblly=6.0cm,
bburx=11cm, bbury=16cm, width=9cm, height=9cm, angle=0, clip=0}
\vspace{-15.5cm} \caption{\label{pdecay} The dominant proton decay
mode.}
\end{figure}

Due to the VEVs of the Higgs triplets $\Sigma_a^{}$, the
$\omega^{-2/3}_{}$ component of the leptoquark triplet $\Omega$ will
have a trilinear coupling with the leptoquark singlets
$\delta_{1,2}^{}$, i.e.
\begin{eqnarray}
\label{lqcubic}
\mathcal{L}&\supset&-\rho\delta^{+1/3}_{1}\delta^{+1/3}_{2}\omega^{-2/3}_{}+\textrm{H.c.}~~\textrm{with}\nonumber\\
&&~~\rho=\kappa_1^{}\langle\sigma^{0\ast}_1\rangle+\kappa_2^{}\langle\sigma^{0\ast}_2\rangle\,.
\end{eqnarray}
By integrating out the leptoquark scalars, we can obtain a low-scale
effective Lagrangian as below,
\begin{eqnarray}
\mathcal{L}^{\textrm{eff}}_{}&=&\sum_{a\neq
b}^{}\frac{\rho^\ast_{}f^\Omega_{ij}f^{\delta_a^{}}_{mn}h^{\delta_b^{}}_{kl}}{m_\Omega^2
m_{\delta_a^{}}^2 m_{\delta_b^{}}^2}(\bar{\nu}_{L_i^{}}^c
u_{L_j^{}}^{})(\bar{\nu}_{L_m^{}}^c
d_{L_n^{}}^{})(\bar{e}_{R_k^{}}^c
u_{R_l^{}}^{})\nonumber\\
&&+\textrm{H.c.}\,.
\end{eqnarray}
The dominant proton decay thus should be
\begin{eqnarray}
p&\rightarrow& e_R^{+}(\mu_R^{+})+\nu_{L_i^{}}^c+\nu_{L_j^{}}^c\,,
\end{eqnarray}
as shown in Fig. \ref{pdecay}. Clearly, the proton decay violates
the $B-L$ number by two units. We can roughly estimate the proton
decay width by
\begin{eqnarray}
\label{pdw} \Gamma_{p\rightarrow
e^{+}_{}\nu\nu}^{}&=&\sum_{ij}^{}\Gamma_{p\rightarrow
e_R^{+}+\nu_{L_i^{}}^c+\nu_{L_j^{}}^c}^{}\nonumber\\
&\propto&\sum_{a\neq b}\sum_{ij}^{}
\frac{|f_{i1}^{\Omega}|^2_{}|f_{j1}^{\delta_a^{}}|^2_{}|h_{11}^{\delta_b^{}}|^2_{}
|\rho|^2_{}}{m_\Omega^4 m_{\delta_a^{}}^4
m_{\delta_b^{}}^4}\,,\nonumber\\
\Gamma_{p\rightarrow
\mu^{+}_{}\nu\nu}^{}&=&\sum_{ij}^{}\Gamma_{p\rightarrow
e_R^{+}+\nu_{L_i^{}}^c+\nu_{L_j^{}}^c}^{}\nonumber\\
&\propto&\sum_{a\neq b}\sum_{ij}^{}
\frac{|f_{i1}^{\Omega}|^2_{}|f_{j1}^{\delta_a^{}}|^2_{}|h_{21}^{\delta_b^{}}|^2_{}
|\rho|^2_{}}{m_\Omega^4 m_{\delta_a^{}}^4 m_{\delta_b^{}}^4}\,.
\end{eqnarray}

\section{Parameter choice}

We now give an example of the parameter choice to show that our
model can simultaneously generate a desired baryon asymmetry and an
testable proton decay. We take
\begin{eqnarray}
&&M_{\Sigma_1}^{}=0.2\,M_{\Sigma_2}^{}=10^{12}_{}\,\textrm{GeV}\,,\nonumber\\
&&
|\mu_1^{}|=0.2\,|\mu_2^{}|=3\times 10^{9}_{}\,\textrm{GeV}\,,\nonumber\\
&&|\kappa_1^{}|=|\kappa_2^{}|=0.1\,,\nonumber\\
&&\sin\left(\frac{\kappa_1^\ast \kappa_2^{}\mu_1^\ast \mu_2^{}
}{|\kappa_1^{} \kappa_2^{}\mu_1^{} \mu_2^{} |}\right)=-0.32\,,
\end{eqnarray}
to derive the out-of-equilibrium quantity
\begin{eqnarray}
K_{1}^{}\simeq 0.62\,,
\end{eqnarray}
and the CP asymmetry
\begin{eqnarray}
\varepsilon_1^{}\simeq\varepsilon_1^{\Omega}\simeq 1.2\times
10^{-8}_{}\gg \varepsilon_1^{l_L^{}}\,.
\end{eqnarray}
The final baryon asymmetry then would be
\begin{eqnarray}
\eta_B^{}=7.04\times \frac{n_B^{}}{s}\simeq 7.04\times3\times
\frac{28}{79}\frac{\varepsilon_1^{}}{g_\ast^{}}\simeq 6.57\times
10^{-10}_{}\,,
\end{eqnarray}
which is consistent with the cosmological observations
\cite{komatsu2011}.

From the above parameter choice, we can also read the trilinear
coupling among the leptoquarks,
\begin{eqnarray}
|\rho|\simeq 1.1\,\textrm{eV}\,.
\end{eqnarray}
Such a tiny parameter means the proton decay (\ref{pdw}) can
naturally have a life time close to the experimental limits
\cite{nakamura2010} $\tau_{p\rightarrow e^{+}_{}\nu\nu}^{}>1.7\times
10^{31}_{}\,\textrm{yr}$ and $\tau_{p\rightarrow
\mu^{+}_{}\nu\nu}^{}
>2.1\times 10^{31}_{}\,\textrm{yr}$ even if the leptoquarks are at the TeV scale.

\section{Summary}

In this paper, we have shown that the baryon asymmetry and the
proton decay can have a common origin. In our model, the decays of
the heavy Higgs triplets can produce a $B-L$ asymmetry in the
TeV-scale leptoquarks. Due to the sphalerons, we eventually can
obtain a baryon asymmetry to explain the baryon asymmetry in the
universe. Benefited from the seesaw-suppressed VEVs of the Higgs
triplets, the leptoquarks can have a trilinear coupling to mediate
an observable proton decay.

\textbf{Acknowledgement}: PHG is supported by the Alexander von
Humboldt Foundation. US thanks R. Cowsik for arranging his visit as
the Clark Way Harrison visiting professor.

\end{document}